\documentclass[useAMS]{mn2e}
\usepackage{psfig}
\usepackage{graphicx}

\usepackage{times}

\def\lsim{ \lower .75ex\hbox{$\sim$} \llap{\raise .27ex \hbox{$<$}} }
\def\gsim{ \lower .75ex \hbox{$\sim$} \llap{\raise .27ex \hbox{$>$}} }
\def\fe{{\it Fermi}}

\title[GRBs observed by \fe--LAT] 
{GeV emission from Gamma Ray Bursts: a radiative fireball?}

\author[Ghisellini et al.]
{G. Ghisellini$^1$\thanks{E--mail: gabriele.ghisellini@brera.inaf.it}, 
G. Ghirlanda$^2$, L. Nava$^{1,2}$, A. Celotti$^2$ \\
$^1$INAF -- Osservatorio Astronomico di Brera, via E. Bianchi 46, I--23807
Merate, Italy\\
$^2$ S.I.S.S.A., V. Beirut 2--4, I--34014 Trieste, Italy 
}

\begin{document}



\maketitle

\begin{abstract} 
We study the emission observed at energies $>$100 MeV of 11 Gamma Ray Bursts 
(GRBs) detected by the \fe/Large Area Telescope (LAT) until October 2009. 
The GeV emission has three main properties: 
(i) 
its duration is often
longer than the duration of the softer emission 
detected by the Gamma Burst Monitor (GBM) onboard \fe\
[this confirms earlier results from the 
Energetic Gamma--Ray Experiment Telescope (EGRET)]; 
(ii) its spectrum is consistent with
$F_{\nu}\propto \nu^{-1}$ and does not show strong spectral evolution; 
(iii) for the brightest bursts, the flux detected by the LAT decays as a power law 
with a typical slope: $t^{-1.5}$.  
We argue that the observed $>$0.1 GeV flux can be 
interpreted as afterglow emission shortly following the start of the 
prompt phase emission as seen at smaller frequencies. 
The decay slope is what expected if the fireball emission is produced in the
radiative regime, i.e. all dissipated energy is radiated away.
We also argue that the detectability in the GeV energy range depends on the bulk Lorentz 
factor $\Gamma$ of the bursts, being strongly favoured in the case of large $\Gamma$.
This implies that the fraction of bursts detected at high energies corresponds 
to the fraction of bursts having the largest $\Gamma$. 
The radiative interpretation can help to explain why the  
observed X--ray and optical afterglow energetics are much smaller
than the energetics emitted during the prompt phase, despite the 
fact that the collision with the external medium should be more 
efficient than internal shocks in producing the radiation we see.
\end{abstract}

\section{Introduction}

The {\it Fermi Gamma Ray Space Telescope (Fermi)} has onboard two
instruments: the Large Area Telescope (LAT), sensitive in the
100 MeV -- 100 GeV energy range (and even beyond 100 GeV, for very bright 
sources, Atwood et al. 2009), and the Gamma Bursts Monitor (GBM), especially designed for the 
detection of Gamma Ray Bursts (GRBs), sensitive in the 8 keV -- 40 MeV
energy range (Meegan et al. 2009).
The LAT revealed 12 GRBs above 100 MeV confirming that
GRBs can be sources of very high energy photons and that the fraction
of GRBs that can be detected at these energies is roughly 10 per cent
of those detected by the GBM at lower energies.
It was the EGRET instrument, onboard the {\it Compton Gamma Ray Observatory (CGRO)}
the first to detect GRBs above 100 MeV (Fishman \& Meegan 1995; 
Kaneko et al. 2008), but it is the much better
sensitivity (and reduced dead time) of the LAT to allow us
for the first time to try to understand the origin of this
emission and to answer the question: 
does it belong to the prompt phase or is it afterglow
emission produced by the fireball colliding with the circum--burst
medium? Or has it still another origin?

One of the puzzling features of the high energy emission as revealed by EGRET
was that it was long lasting, yet it started during the prompt
phase as seen by the Burst Alert and Transient Experiment (BATSE)
onboard {\it CGRO} sensitive in the 30 keV -- 1 MeV energy band.
For instance, GRB 940217 emitted $>100$ MeV photons up to 1.5 hours after the
prompt phase ended in the BATSE detector. 
A photon of 18 GeV was received $\sim$5000 s after the trigger (Hurley et al. 1994), 
and this was the highest photon energy of a GRB until the \fe--LAT era.
On the other hand, about a third of the high energy photons were
received within 120 s, before the end of the prompt phase as 
detected by BATSE.

Up to now, there have been three LAT--detected GRBs already discussed 
in the literature. 
In GRB 080916C (Abdo et al. 2009a), there is evidence that the
spectrum from 8 keV to 10 GeV can be described by the same Band
function (i.e. two smoothly connected power laws), suggesting
that the LAT flux has the same origin of the low energy flux.
On the other hand, the flux level of the LAT emission, its spectrum and its
long lasting nature match the expectations from a forward shock,
leading Kumar \& Barniol--Duran (2009) to prefer the ``standard afterglow"
interpretation (see also Razzaque, Dermer \& Finke 2009 for an hadronic model;
Zhang \& Peer 2009 for a magnetically dominated fireball model and
Zou et al. 2009 for a synchrotron self--Compton origin).

\begin{table*}
\caption{
The 12 bursts detected by the \fe--LAT instrument above 100 MeV, 
until October 03 2009.
Besides their redshifts (when measured) and duration, 
we give the parameters of the time integrated GBM spectrum
collected from the literature and the corresponding reference.
Fluences $S$ are in [erg cm$^{-2}$], peak energies $E_{\rm peak}$ in keV.
In column 10 we report the fluence $S$ in the [8 keV--10 MeV] energy 
range calculated from the spectral parameters of the GBM.  
Column 11 reports the fluence in the [0.1--100 GeV] energy 
range obtained from the analysis of the LAT spectra performed in this paper (whose results 
are given in Tab. 2). We adoped a BAND model for the GBM, and a simple power law of photon 
slope $\Gamma$ for the LAT. When $\beta$ is not indicated, the adopted fitting model is a cut off 
power--law of photon slope $\alpha$.
$^a$: $S_{\rm GBM}$ in the [8 keV--30 MeV] energy range
$^b$: $S_{\rm GBM}$ in the [50 --300 keV] energy range;
$^c$: $S_{\rm GBM}$ in the [50 keV--40 MeV] energy range;
$^d$: $S_{\rm GBM}$ in the [50 keV--10 MeV] energy range.
The number quoted in the ``Ref." column refer to GCN circulars as follows:
8141: van der Horst \& Connaughton 2008;
8278: van der Horst \& Goldstein  2008;
8407: Omodei 2008;
8682: Chaplin, van der Horst \& Preece 2008; 
8902: von Kienlin 2009;
9021: Ohno M. et al. 2009;
9057: Rau, Connaughton \& Briggs 2009;
9336: Guiriec, Connaughton \& Briggs 2009;
9579: von Kienlin 2009;
9866: Bissaldi \& Connaughton 2009;
9933: Bissaldi 2009;
9983: Rau 2009.
}
\label{tab0}      
\centering   
\begin{tabular}{l l l l l l l l l l l}      
\hline
\hline       
GRB     &$z$     &$T_{90}$ &$S_{\rm GBM}$ & $\alpha_{\rm GBM}$ 
                           &$\beta_{GBM}$ &$E_{\rm peak}$   &Ref   &$E_{\rm \gamma, iso}$ &$S_{\rm GBM}$  &$S_{\rm LAT}$ \\
        &        &s        &                        &                  &                &keV           &     &erg    &8--10$^4$ keV  &0.1--100 GeV \\             
\hline
080825C &...   &22         &2.4e--5                 &--0.39$\pm$0.04   &--2.34$\pm$0.09 &155$\pm$5     &8141 &...    &(3.4$\pm$0.3)e--5   &(9.5$\pm$4)e--6  \\
080916C &4.35  &66         &1.9e--4$^a$             &--0.91$\pm$0.02   &--2.08$\pm$0.06 &424$\pm$24    &8278 &5.6e54 &(1.6$\pm$0.2)e--4   &(7$\pm$1)e--5  \\
081024B &...   &0.8        &(3.4$\pm$0.1)e--7       &--0.70$\pm$0.13   &...             &1583$\pm$520  &8407 &...    &(3.2$\pm$0.1)e--6   &(3$\pm$2)e--6\\
081215  &...   &$\sim$90   &(2.8$\pm$0.5)e--6$^b$   &--0.14$\pm$0.26   &...             &139$\pm$14    &8682 &...    &                    &... \\
090217  &...   &32.8       &(3.08$\pm$0.03)e--05    &--0.845$\pm$0.023 &--2.86$\pm$0.3  &610$\pm$32    &8902 &...    &(3.8$\pm$0.4)e--5   &(4.2$\pm$1.6)e--6\\
090323  &3.57  &$\sim$150  &(1.00$\pm$0.01)e--4     &--0.89$\pm$0.03   &...             &697$\pm$51    &9021 &3.4e54 &(1.32$\pm$0.03)e--4 &(3.6$\pm$0.8)e--5 \\
090328  &0.736 &$\sim$25   &(8.09$\pm$0.10)e--5     &--0.93$\pm$0.02   &--2.2$\pm$0.1   &653$\pm$45    &9057 &2.1e53 &(1.52$\pm$0.02)e--4 &(3.3$\pm$2)e--5 \\
090510  &0.903 &1          &(3.0$\pm$0.2)e--5$^c$   &--0.80$\pm$0.03   &--2.6$\pm$0.3   &4400$\pm$400  &9336 &5.0e52 &(2.3$\pm$0.2)e--5   &(3.7$\pm$0.7)e--5\\
090626  &...   &70         &(3.5$\pm$0.1)e--5       &--1.2$\pm$0.02    &--1.98$\pm$0.02 &175$\pm$12    &9579 &...    &(6.0$\pm$0.2)e--5   &(9.6$\pm$6)e--6\\
090902B &1.822 &$\sim$21   &(3.74$\pm$0.03)e--4$^d$ &--0.696$\pm$0.012 &--3.85$\pm$0.25 &775$\pm$11    &9866 &4.4e54 &(5.4$\pm$0.04)e--4  &(5.9$\pm$0.6)e--4\\
090926A &2.106 &20$\pm$2   &(1.45$\pm$0.04)e--4     &--0.75$\pm$0.01   &--2.59$\pm$0.05 &314$\pm$4     &9933 &2e54   &(1.9$\pm$0.05)e--4  &(4.3$\pm$0.8)e--5\\
091003  &0.897 &21$\pm$0.5 & (3.76$\pm$0.04)e--5    &--1.13$\pm$0.01   &--2.64$\pm$0.24 &86.2$\pm$23.6 &9983 &8.7e52 &(4.16$\pm$0.03)e--5 &(1.3$\pm$0.8)e--5 \\
\hline                  
\hline                  
\end{tabular}
\end{table*}

\begin{figure*}
\hskip 0.3 cm
\psfig{figure=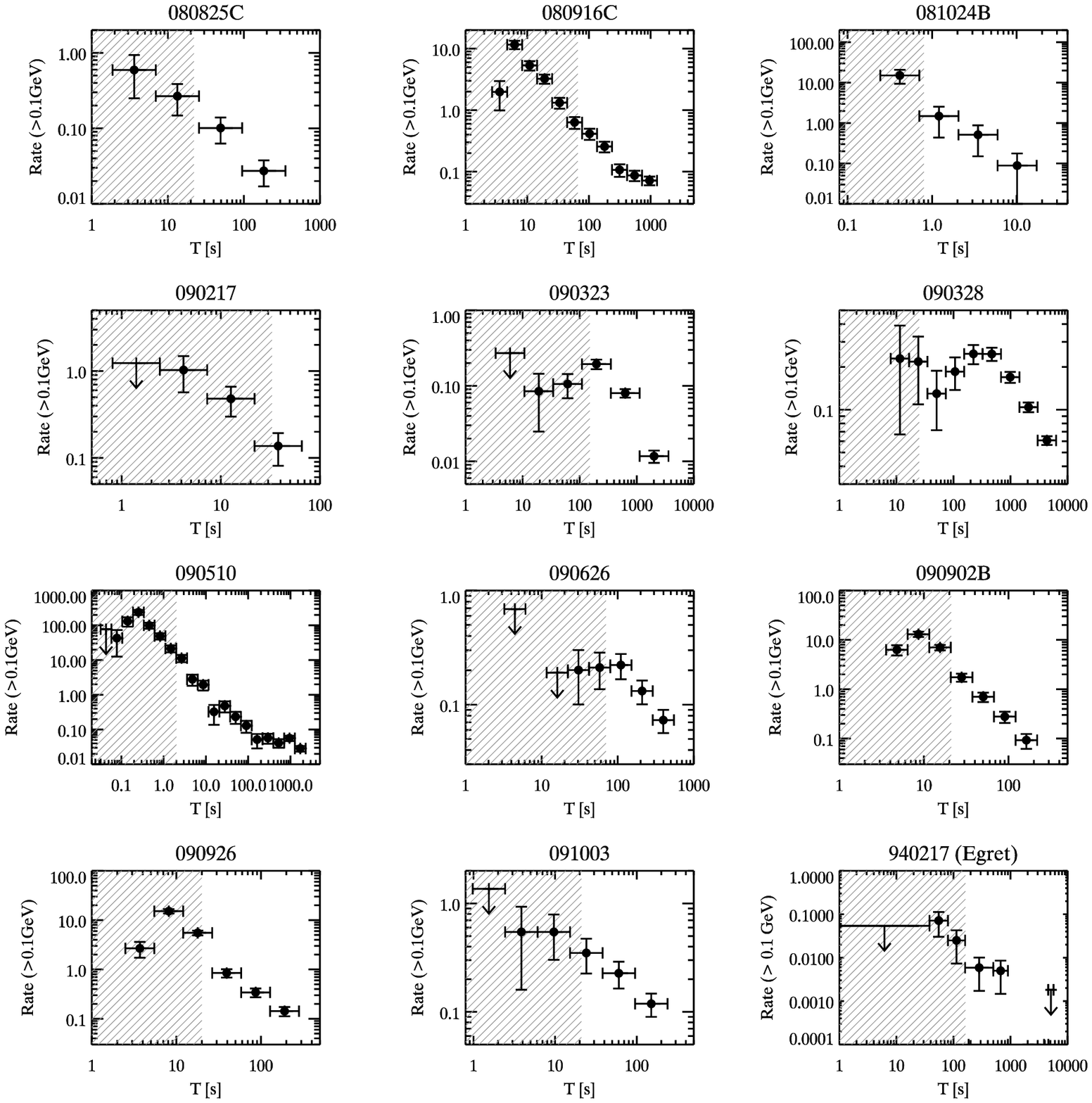,width=18cm} 
\vskip -0.2 cm
\caption{
Light curves of the 11 GRBs detected by LAT 
plus GRB 940217, as detected by EGRET
(bottom right panel).
The hatched region represents the duration
($T_{90}$) of the emission detected by the GBM in the 8 keV--40 MeV energy range
(for GRB 940217 it refers to the emission detected by BATSE).
Times are in the observer frame for all bursts and arrows represent 2$\sigma$ upper limits.}
\label{lc}
\end{figure*}

In the short bursts GRB 090510 the spectrum in the LAT energy range 
is not the extrapolation of the flux from lower energies, but is harder,
leading  Abdo et al. (2009b) to propose a synchrotron self--Compton interpretation
for its origin.
Instead we (Ghirlanda, Ghisellini \& Nava 2009) proposed that the LAT flux
is afterglow synchrotron emission, on the basis of its time profile and
spectrum (see also Gao et al. 2009; De Pasquale et al. 2009).

Finally, the LAT flux of GRB 090902B decays as $t^{-1.5}$ (Abdo et al. 2009c),
it lasts longer than the flux detected by the GBM, and its spectrum is harder than the 
extrapolation from lower frequencies, making it a good candidate for an afterglow
interpretation, despite the arguments against put forward by Abdo et al. (2009c),
that we will discuss in this paper. 
Moreover, in GRB 090902B there is evidence of a soft excess 
(observed in the GBM spectrum below 50 keV) which is spectrally
consistent with the extrapolation at these energies of the LAT spectrum.

As the few examples above demonstrate, there is no consensus yet
on the nature of the high energy emission of GRBs.
Since only three of the nearly dozen bursts detected by the LAT have already
been discussed in the literature, we present here a study of the entire
sample of bursts detected at high energies by the LAT.
We will construct the light curves of the high energy flux and 
the spectral shape in the 0.1--100 GeV energy range,
to find if there are properties that are common among
different bursts that can help to understand their nature.

Indeed, we believe that a consistent scenario emerges: the LAT spectra 
are often inconsistent with the extrapolation of the GBM spectra
(except two cases) and the light curves can be often described
by a power law decay in time, i.e. $F_{\rm LAT}\propto t^{-\alpha}$,
with a slope often close to $\alpha=1.5$.
In the brightest cases also the rising part is visible, and
is consistent with $F_{\rm LAT}\propto t^2$.
These are, in our opinion, strong indications of the afterglow
nature of the LAT emission.
Furthermore, we suggest that GRBs with a flux 
decaying as $F_{\rm LAT}\propto t^{-1.5}$, and with a 
spectral slope around unity [i.e. $F(\nu)\propto \nu^{-1}$],
are emitting in the radiative regime of a forward shock.
We will also point out the role that the electron--positron pair
production process has in establishing the radiative regime.
Finally, we will discuss the consequences of our findings.

We adopt a cosmology with $h=\Omega_\Lambda=0.7$ and $\Omega_{\rm M}=0.3$
and the convention $Q = 10^x Q_x$, using cgs units.

\section{Sample and data analysis}

We considered all the 12 bursts detected in the \fe--LAT until the 3rd of October 2009. 
These are reported in Tab. 1 with their redshifts (Col. 2) and the 
spectral parameters and fluences (Col. 4, 5, 6, 7) as reported in the literature, 
obtained from the analysis of the GBM spectrum. 
Since the GBM fluences reported in the literature refer to different energy ranges,
we convert all the GBM fluences to the common 8 keV -- 10 MeV energy range (Col. 10). 
In addition (last column) we report the fluences in 
the 0.1--100 GeV energy range of the LAT obtained from the spectral analysis of the 
LAT data (spectral
parameters are given in Tab. 2). For those GRBs with measured redshifts we computed the isotropic
equivalent energy $E_{\gamma,\rm iso}$ by integrating the GBM spectral model in the 
in the 1 keV -- 10 MeV rest frame energy range.

Among the considered bursts there are three cases which have been published in recent papers:
GRB 080916C (Abdo et al. 2009a), 
GRB 090510 (Abdo et al. 2009b; Ghirlanda et al. 2009) and 
GRB 090902B (Abdo et al. 2009c). 
All the others are unpublished.  
We did not consider GRB 081215 which, lying at a large angle (86$^{\circ}$) 
with respect to the  LAT boresight (Preece et al. 2008), required a non 
standard analysis of the LAT data (McEnery et al. 2008). 
GRB 081024B and GRB 090510 are of the short class. 
Seven bursts have measured redshifts, for all the others we assume a typical 
redshift of 2 and 1 for the long and short class.

We have analysed the \fe--LAT data\footnote{http://fermi.gsfc.nasa.gov/ssc/data/} 
with the  \fe\  \texttt{ScienceTools} \texttt{(v9r15p2)} 
released on Aug. 8th 2009.  
LAT count light curves (extracted with the  \texttt{gtbin} tool) were rebinned in time with 
a variable bin size, different for each burst. 

We analysed the spectrum of the emission detected by the LAT. For the brightest part of the 
burst we applied the standard procedure (i.e. extracted the spectra 
and created the relative response files with the \texttt{gtbin} and \texttt{gtrspgen} 
tools, respectively).  
We considered the spectrum over a time interval covering
entire light curve, and if the burst was particularly bright we also extracted the 
spectrum over a time interval coincident with the duration of the emission as observed 
by the GBM. 
To verify if and at what extent the LAT spectrum could vary with time, we 
extracted a series of consecutive spectra for each burst. 
As in most bursts we 
did not find evidence for substantial spectral evolution of the LAT component, we used 
the average spectrum to convert the count rate into physical  units.

\section{Results}

{\bf Light curves --} 
Fig. \ref{lc} shows the light curves obtained from the selection of 
the LAT events with energies $>0.1$ GeV. 
In each plot we also show the time interval (hatched region) 
corresponding to the duration of the GBM light curve ($T_{90}$ in Tab. 1). 
In 9/11 events there is 
a peak in the LAT light curve and the latter has a duration much longer than
the duration of the GBM light curve (shown by the hatched region in Fig. \ref{lc}). 
After the peak, the light curves of different GRBs show a similar temporal decay. 
In a few cases (see also Ghirlanda et al. 2009) a rising of the light curve as 
$t^2$ is seen before the peak. 
The three faintest GRBs (GRB 090323, GRB 090328 and GRB 090626)
have light--curves that appear much flatter than the other ones
(please note the different scale of their $y$--axis) and we cannot exclude
that the background, in this cases, plays some role.
The bottom right panel shows the light--curve of GRB 940217
as detected by EGRET (Hurley et al. 1994), 
selecting photons above 100 MeV.
As can be seen, also this burst show a similar decaying light curve.

\vskip 0.3 cm
\noindent
{\bf Spectral evolution --}
In Tab. 2 we report the results of the LAT spectral analysis. For each burst the first 
line refers to the spectrum used to convert the count rate into 
physical units while the following
lines give the spectral index for each time resolved spectrum. 
We report in Tab. 2 also the flux integrated between 100 MeV and 100 GeV. 
By comparing the time resolved spectral results of individual bursts we see 
that there is no evidence of strong spectral evolution of the LAT spectral index during the burst. 
On average, all the spectral index are distributed between 1.5 and 2.2. 

\begin{figure}
\vskip -0.5 cm
\hskip -1.5cm
\psfig{figure=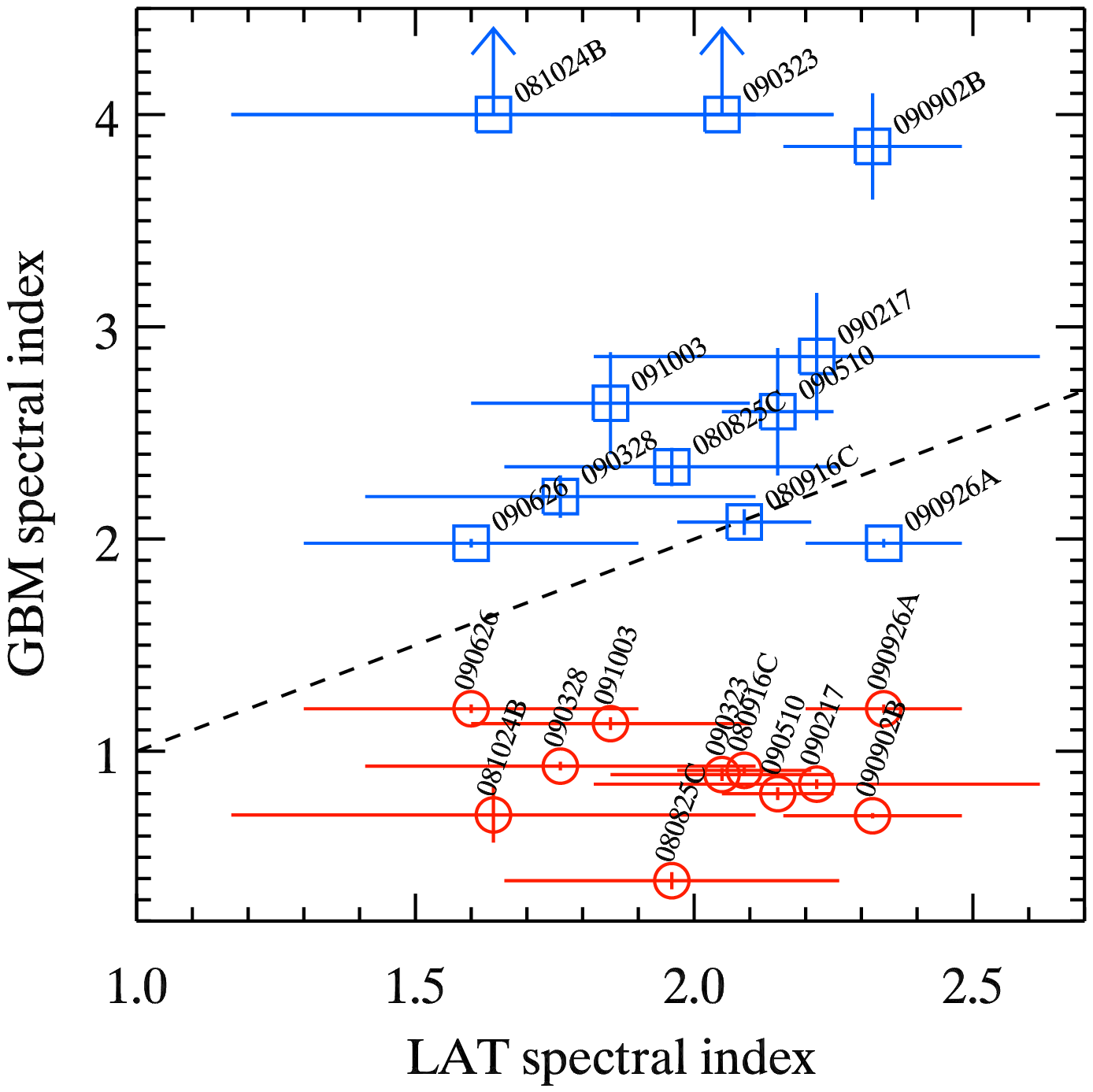,width=10.5cm}
\vskip -0.2 cm
\caption{
Spectral index ($\alpha$ and $\beta$, red circles and blue squares respectively) 
of the GBM time integrated spectra (reported in 
Tab. 1) versus the spectral index obtained from the analysis 
of the LAT data presented in this paper (Tab. 2). 
The dashed line represents equality. 
The two lower limits
are GRB 081024B and GRB 090323 whose time integrated GBM spectrum is best fit by a 
cutoff power--law model. For illustrative purposes we assumed 
$\beta>4$ for these two bursts.  This plot shows that the LAT spectrum is 
softer than 
the low energy spectral index of the Band model fitting the GBM spectrum 
(red circles) 
and it is harder than the high energy spectrum of the Band model fitting the GBM 
spectrum (blue squares). 
}
\label{galbe}
\end{figure}

\vskip 0.3 cm
\noindent
{\bf Spectral slopes in the LAT vs GBM --}
In Fig. \ref{galbe} we compare the spectral index of the LAT emission with the spectral index 
of the average GBM spectrum (whose spectral parameters are reported in Tab. 1). 
The low energy spectral index $\alpha$ 
(circles in Fig. \ref{galbe}, red in the electronic version) of the Band model 
(or of the cutoff power--law model for GRB 081024B and GRB 090323) 
is systematically harder than the spectral index of the LAT component.
The high energy spectral index $\beta$ of the Band model (open squares in 
Fig. \ref{galbe}, blue in the electronic version) appears softer than the LAT spectrum. 
An extreme case is GRB 090902B which clearly shows that the LAT component is 
spectrally different from the tail of the Band function. 
Indeed, in this burst there is also evidence of a soft spectral 
excess detected in the GBM below 50 keV (Abdo et al. 2009c; De Palma et al. 2009). 
We also note that in only two bursts, GRB 080916C (Abdo et al. 2008) and GRB 090926 
the high energy spectrum of the Band model is consistent with the spectral slope 
of the LAT data. 

\begin{table}
\caption{LAT spectral results. We give the time interval ($t_0-t_1$) 
for the accumulation of each spectrum, the photon index, the flux integrated between 
100 MeV and 100 GeV, and the C--statistic for degrees of freedom. 
Errors are given at the 90\% 
confidence level. }
\label{tab1}      
\centering  
\begin{tabular}{llllll}     
\hline
\hline       
GRB &$t_0$  &$t_1$      &$\Gamma_{\rm LAT}$  &C$_{\rm stat}$/dof	&Flux \\
    &s      &s  	&          &                    &erg cm$^{-2}$s$^{-1}$ \\
\hline
080825C	&0    &200  &1.96$\pm$0.3  &10/6	&(4.0$\pm$2.0)e--8	\\
080916C	&0    &200  &2.09$\pm$0.12 &19/15	&(3.3$\pm$0.7)e--7        \\
081024B &0    &5    &1.64$\pm$0.47 &5.6/5	&(4.0$\pm$3.0)e--7	\\
...	&0    &1    &2.0$\pm$0.7   &		&			\\
...	&1    &5    &1.65$\pm$0.8  &		&			\\
090217 	&0    &100  &2.22$\pm$0.4  &4/5		&(4.0$\pm$3.3)e--7	\\
090323 	&0    &400  &2.05$\pm$0.2  &6/10	&(7.9$\pm$4.0)e--7    \\
...	&0    &200  &2.16$\pm$0.3  &		&		      \\
...	&200  &400  &1.98$\pm$0.23 & 		&			\\
090328 	& 0   &100  &1.76$\pm$0.35 &8.8/10	&(1.2$\pm$0.2)e--7	\\
...	&100  &200  &1.61$\pm$0.23 &		&			\\
...	&200  &400  &1.81$\pm$0.25 &		&			\\
090510 	&0    &7    &2.15$\pm$0.1  &23/30	&(4.7$\pm$1.0)e--6    \\
...	&0.1  &0.324&1.8$\pm$0.25  &		&			\\
...	&0.324&1.05 &2.28$\pm$0.23 &		&		     \\
...	&1.05 &6.12 &2.22$\pm$0.28 &            &		     \\
090626 	&0    &600   &1.7$\pm$0.12 &8/10	&(4.7$\pm$1.0)e--8  \\
...  	&0    &70   &1.6$\pm$0.3   &    	&  \\
...	&70   &170  &1.99$\pm$0.33 & 		&		     \\
...	&170  &600  &1.65$\pm$0.3  &		&			\\
090902B	&0    &320  &2.32$\pm$0.16 &6/10	&(1.8$\pm$0.3)e--6    \\
...	&4    &6    &2.67$\pm$0.64 &		&		     \\	
...	&6    &9    &2.34$\pm$0.51 &		&		      \\	
...	&9    &10.5 &2.5$\pm$0.43  &		&			\\	
...	&10.5 &12.5 &2.37$\pm$0.47 &		&			\\	
...	&12.5 &21   &1.92$\pm$0.27 &		&			\\	
...	&21   &40   &1.76$\pm$0.28 &		&			\\	
...	&40   &80   &1.84$\pm$0.3  &		&		       \\	
...	&80   &160  &1.73$\pm$0.52 &		&			\\	
...	&160  &320  &1.91$\pm$0.44 &		&			\\	
090926A	&0    &25   &2.34$\pm$0.14 &4/10	&(1.7$\pm$0.3)e--6     \\
...	&2    &8    &2.75$\pm$0.5  &		&			\\
...	&8    &15   &2.36$\pm$0.22 &		&			\\
...	&15   &25   &2.0$\pm$0.23  &		&			\\
...	&25   &75   &1.85$\pm$0.22 &		&			\\
...	&75   &225  &2.09$\pm$0.42 &		&		        \\
091003	&0    &100  &1.85$\pm$0.25 &12/7	&(7.4$\pm$1.2)e--8     \\
...	&100  &200  &1.81$\pm$0.4  &		&			\\
...	&200  &400  &1.8$\pm$0.2   &		&			\\
\hline
\hline                  
\end{tabular}
\end{table}

\begin{figure}
\vskip -0.5 cm
\hskip -0.8 cm
\psfig{figure=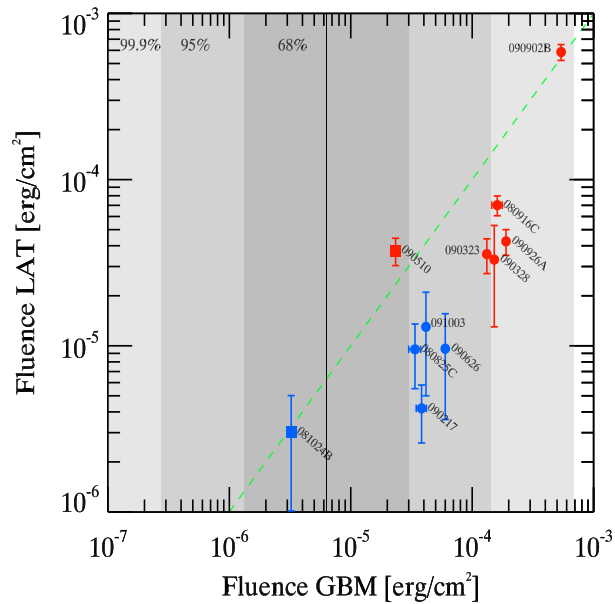,width=9.5cm,height=9cm}
\vskip -0.2 cm
\caption{
Fluence in the [0.1--100 GeV] LAT energy range as a function of the
[8 keV -- 10 MeV] GBM ones.
Short GRBs are marked with filled squares, long GRBs with filled circles.
GRBs with known redshifts are the ones with a LAT fluence
larger than $2\times 10^{-5}$ erg cm$^{-2}$ (red in the electronic version).
The shaded areas indicate the 1--2--3 $\sigma$ values
of the distribution of fluences of the 121 GRB with $E_{\rm peak}$ (as of Oct. 2009) 
detected by the GBM.}
\label{fluence}
\end{figure}

\vskip 0.3 cm
\noindent
{\bf LAT vs GBM fluences --}
Fig. \ref{fluence} shows the fluence in the LAT energy range 100 MeV -- 100 GeV 
(using the fluxes listed in Tab. 2) 
as a function of the fluence in the GBM energy range 8 keV -- 10 MeV.
The shaded regions correspond to 1, 2 and 3 $\sigma$ of the distribution of GBM
fluences for the 121 GRBs detected so far by the GBM with measured prompt phase
emission peak energy (Nava et al., in preparation) and that appeared in the 
Gamma Ray Bursts Coordinate Network (GCN) circulars\footnote{http://gcn.gsfc.nasa.gov/}.
The dashed line marks equality between the two fluences.
We can see that all but the two short bursts (GRB 081024B and GRB 090510)
have GBM fluences much brighter than average.
If all GRBs with GBM fluences 1$\sigma$ brighter than average and
in the LAT field of view (i.e. one half) were detected by the LAT, 
we should have a fraction of LAT--detected GRBs of 
$\sim 16\%$, that is not far from the actual fraction
(see also Guetta \& Pian 2009).
One can compare Fig. \ref{fluence} with Fig. 4 of Le \& Dermer (2009),
showing the pre--{\it Fermi} bursts detected by EGRET and BATSE.
Apart from GRB 930131, showing an EGRET fluence comparable to the BATSE one,
all the other pre--{\it Fermi} GRBs seem to be characterized by a fainter
GeV fluence relative to their fluence at smaller energies, but the sample
is too small to draw any conclusion.

\begin{figure}
\vskip -0.5 cm
\hskip -0.9cm
\psfig{figure=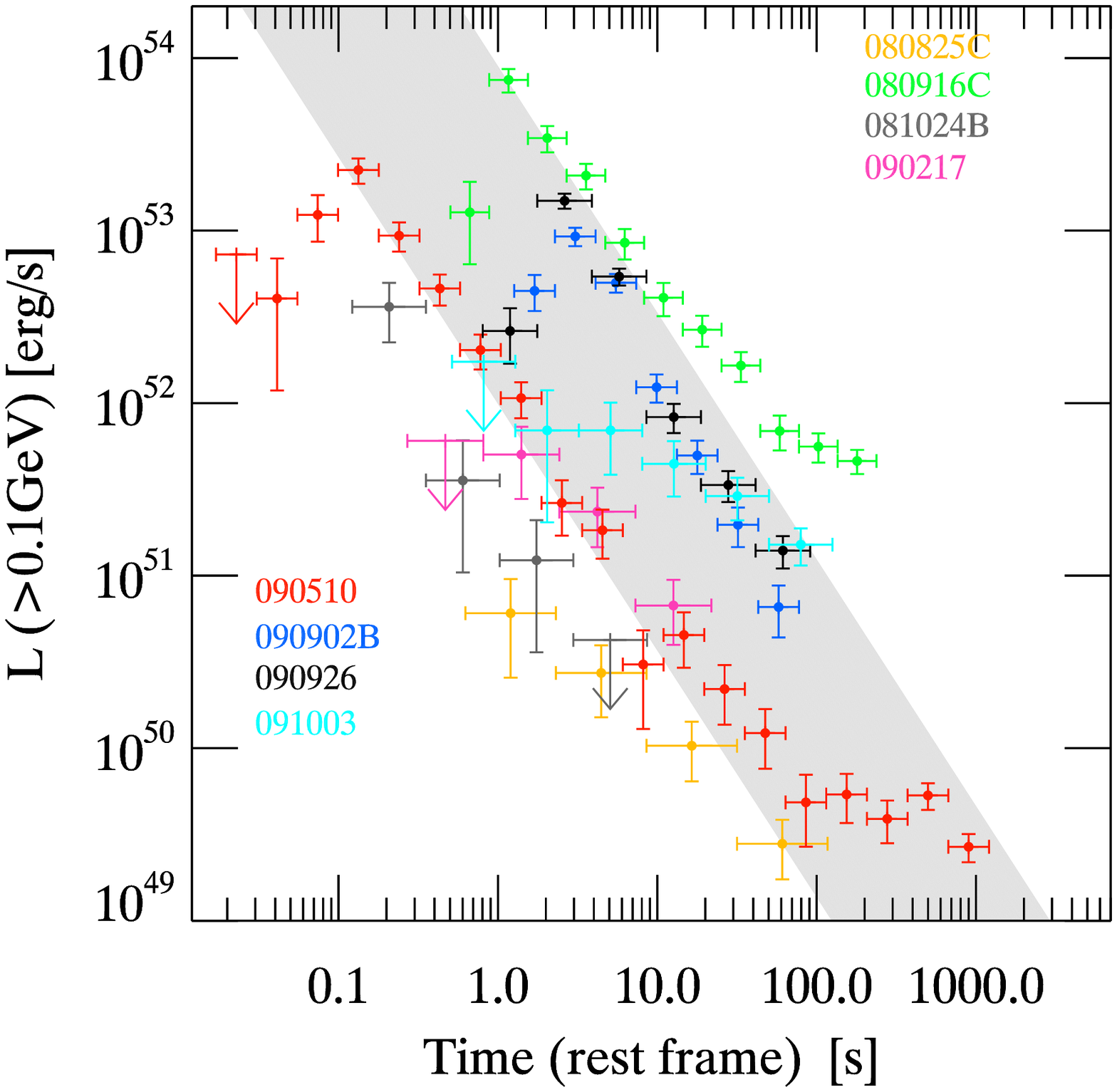,width=9.9cm}
\vskip -0.5 cm
\psfig{figure=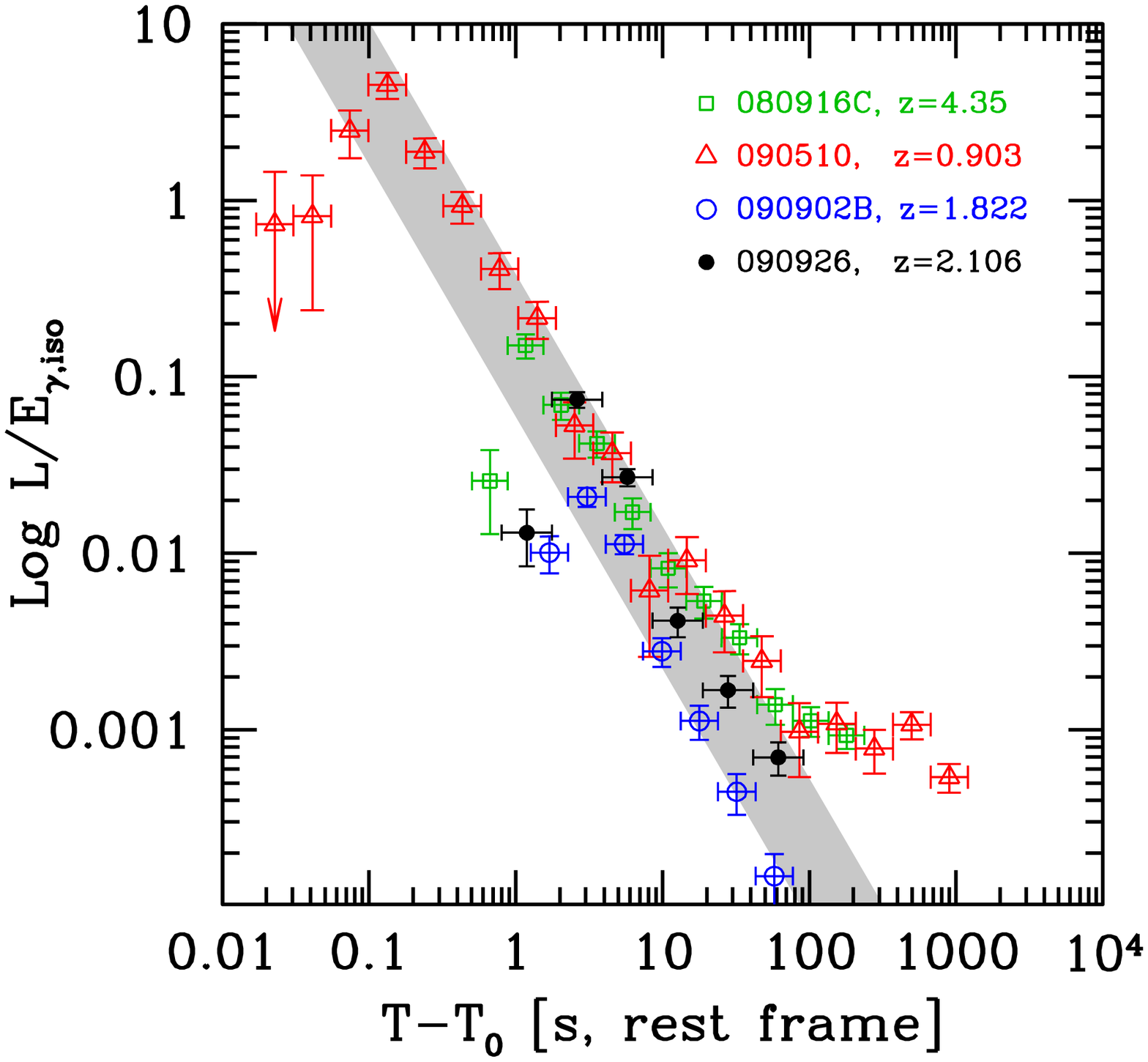,width=9.cm}
\vskip -0.5 cm
\caption{Top panel:
Light curves of the 8 brightest bursts GRBs detected by LAT. 
The luminosities are integrated in the 0.1--100 GeV energy range at the source rest frame. 
For GRBs without measured redshifts we  assumed $z=1$ for short and $z=2$ for 
long events. 
The time is in the rest frame of the sources. 
Upper limits are at 2$\sigma$ level. 
The grey stripe indicates a slope t$^{-10/7}$.
Bottom panel:
Light curves of the 4 brightest GRBs with redshift,
normalised to the total energetics of the GBM energetics.
The luminosities are 
integrated in the 100 MeV--100 GeV energy range at the source rest frame. 
For GRBs without measured redshifts it is assumes 1 for short and 2 for long events. 
The time is in the rest frame of the sources. 
Upper limits are at 2$\sigma$ level. 
The yellow stripe indicates a slope t$^{-10/7}$. 
}
\label{f2}
\end{figure}

\vskip 0.3 cm
\noindent
{\bf Time decay of the LAT flux --}
We converted the count rate of Fig. \ref{lc} into luminosity. For the bursts without 
measured redshifts we assumed a typical redshift of 2 for long events, while for 
GRB 081024B we used a redshift of 1. 
We show the light curves of 8 GRBs with good quality data
in the top panel of Fig. \ref{f2}, where the times are in the source rest frame. 
The grey shaded stripe has a slope of $t^{-10/7}$, and it is shown for comparison.
We can see that the light--curves show a power--law behaviour, 
and that the decay slope is often steeper than unity.
Initially, some bursts show a rising phase and therefore it is possible
to define the peak time of their high energy emission.
As seen below, if the peak time marks the onset of the afterglow emission
it can be used to estimate the bulk Lorentz factor $\Gamma$.

\vskip 0.3 cm
\noindent
{\bf Common decay for the brightest LAT bursts --}
The bottom panel of Fig. \ref{f2} shows the light curves of the 4 brightest GRBs with
redshift, once the 0.1--100 GeV luminosity is divided by 
the energetics $E_{\rm \gamma, iso}$ of the flux detected by the GBM.
The shaded stripe has a slope $t^{-10/7}$, and it is shown for comparison.
These four GRBs show a common behaviour, being all consistent,
within the errors, with the same decay, both in slope and in normalisation.
Note that GRB 090510, a short bursts, behaves similarly to the other 3 bursts,
that belong to the long class, but its light--curve begins much earlier.
If we divide the light--curves by the average luminosities
as derived by the GBM [instead of the energetics; i.e. we multiply by 
the time $T_{90}/(1+z)$] the resulting light--curves of the 4 GRBs spread within a 
larger region.

\vskip 0.3 cm
To conclude, we find that 
i) the  LAT fluxes decay as a power--laws;
ii) the spectral shape at high energies is not strongly evolving;
iii) the LAT spectrum has a slope intermediate between the low and high energy slope 
(i.e. $\alpha$ and $\beta$) of the Band function used to fit the GBM data;
iv) the brightest 4 GRBs show a common $t^{-1.5}$ decay and even the same
normalisation, once their LAT luminosities are divided by the GBM energetics.

These characteristics are the same as observed/predicted by the 
external shock scenario giving rise to the afterglow.
We therefore suggest that the high energy emission of the
GRBs detected by the LAT has an afterglow origin.
The fact that the high energy emission overlaps in time with the 
prompt phase as seen in the GBM can be explained by invoking a 
relatively large value of the bulk Lorentz factor,
corresponding to relatively small deceleration radii and
onset times largely contracted by the Doppler effect.
What is at odd with respect to the ``standard afterglow" scenario
is the relatively steep slope of the flux decay, even when 
the high energy spectrum indicates that we are observing this
component close to its spectral peak.
We offer a solution to this problem in the next section, where we
will also argue that the likely emission process producing the
high energy flux is synchrotron radiation.

\section{The bolometric afterglow luminosity}

In the early afterglow phases, the emission
is likely to occur in the {\it fast cooling} regime,
in which all the energy of the accelerated electrons
is radiated away. 
In this case the bolometric afterglow luminosity
can be calculated in a simple way.
Assume that the shock generated by the fireball
has reached a radius $R$, and that it moves 
within a region characterised by a uniform
number density $n$ (this case can be easily generalised 
to different density radial profiles).
The (comoving) emitting volume is $V^\prime =4\pi R^2 \Delta R^\prime$,
since we are assuming that the fireball is a spherical shell.
The radiative cooling rate of the electrons 
is measured by $\dot \gamma$ where $\gamma m_{\rm e} c^2$ is
the electron energy, and the emitting particles are distributed 
in energy according to $N(\gamma)$.
Note that the time derivative, the electron energies and their
energy distribution are all measured in the comoving frame.
In this case the bolometric luminosity is:
\begin{eqnarray}
L_{\rm iso}\, &=& \, 
\Gamma^2 m_{\rm e}c^2 \int V^\prime N(\gamma)\dot\gamma d\gamma
\nonumber  \\
&=&\, 4 \pi R^2 \Gamma^2 m_{\rm e}c^2  \int N(\gamma)
\dot\gamma \Delta R^\prime(\gamma) d\gamma \nonumber  \\
 &=&\, 4 \pi R^2 \Gamma^2 m_{\rm e}c^3  \int N(\gamma)\gamma  d\gamma
\label{liso1}
\end{eqnarray}
We have used the fact that the distance $\Delta R^\prime$ can be 
approximated by the cooling length as measured in the comoving frame:
$\Delta R^\prime(\gamma)= ct^\prime_{\rm cool} =c \gamma/\dot\gamma$.
Therefore $\Delta R^\prime$ is energy dependent, it is smaller for
high energy particles, that spend most of their energy faster.
Eq. \ref{liso1} is remarkably {\it independent of the specific
radiation process.}
The integral in Eq. \ref{liso1} must correspond to the
fraction $\epsilon_{\rm e}$ of the available energy density
as measured in the comoving frame, i.e.
\begin{equation}
m_{\rm e}c^2 \int N(\gamma)\gamma  d\gamma\, =\,
\epsilon_{\rm e} n \Gamma^2 m_{\rm p} c^2.
\label{epse}
\end{equation}
Therefore Eq. \ref{liso1} becomes:
\begin{eqnarray}
L_{\rm iso} \, &=&\, 4 \pi R^2 \Gamma^4m_{\rm p} c^3  \epsilon_{\rm e} n  
\nonumber  \\
&=&\, 16 \pi a^2 t^2 \Gamma^8  m_{\rm p} c^5 \epsilon_{\rm e} n 
\label{liso2}
\end{eqnarray}
where we have assumed that the size $R$ is measured by the observed time as
$R = 2a ct\Gamma^2$. 
The factor $a$ is equal to 1 if the fireball moves at a constant speed,
and becomes greater than 1 when it decelerates (see e.g. Sari 1997). 
Eq. \ref{liso2} is valid as long as the afterglow is in the fast cooling
regime, irrespective of the radiative or adiabatic nature of the process,
that changes only the relation between the observed time $t$ and the bulk
Lorentz factor $\Gamma$ at that time.
In fact, when the forward shock is coasting (i.e. before being
notably decelerated) we have $L_{\rm iso}\propto t^2$ in both cases.
When the shock starts to decelerate, the observed luminosity decreases 
according to the appropriate $\Gamma(t)$ function, that is different
for the adiabatic and radiative cases.

\vskip 0.3 cm
\noindent
{\bf Adiabatic case --}
We adopt the following relation between the observed time and $\Gamma$:
\begin{equation}
\Gamma^8 \, =\, { 3E_{\rm k,f} \over 32 \pi a^3 n m_{\rm p} c^5 t^3}
\label{tadia}
\end{equation}
where $E_{\rm k,f}$ is the kinetic energy of the fireball after
the prompt phase.
The same equation can be used to define the deceleration time $t_{\rm dec}$,
once we set $a=1$ and substitute $\Gamma_0$ to $\Gamma$.
If $\eta$ is the efficiency of conversion of the initial kinetic 
energy $E_{\rm k,0}$
into radiation of the prompt phase, we have 
\begin{equation}
E_{\rm k,f} \, =\, E_{\rm k,0}-E_{\rm \gamma,iso} \, =\, 
E_{\rm \gamma,iso}\, \left( {1-\eta \over \eta} \right)
\label{eta}
\end{equation}
When the fireball is still in its coasting phase the observed
luminosity increases as $t^2$ due to the increased visible area.
After $t_{\rm dec}$ the observed luminosity decreases as $t^{-1}$,
as can be seen inserting Eq. \ref{tadia} into Eq. \ref{liso2}:
\begin{eqnarray}
L_{\rm iso,a} \, &=&\, 16 \pi  t^2 \Gamma_0^8  m_{\rm p} c^5 
\epsilon_{\rm e} n; \quad t\ll t_{\rm peak} 
\nonumber  \\
L_{\rm iso,a} \, &=&\, 
{3 \over 2 a} \, {\epsilon_{\rm e} E_{\rm k,f} \over t};  
\quad\quad\quad \quad \quad t\gg t_{\rm peak} 
\nonumber  \\
t_{\rm peak,a} \, &=&\, \left[ { 3  E_{\rm k,f} \over
32\pi a \Gamma_0^8 n m_{\rm p} c^5}\right]^{1/3} \, =\,  
{ t_{\rm dec}\over a^{1/3} }
\nonumber  \\
t_{\rm dec} \, &\equiv&\, \left[ { 3  E_{\rm k,f} \over
32\pi \Gamma_0^8 n m_{\rm p} c^5}\right]^{1/3} 
\label{liso3}
\end{eqnarray}
To find $t_{\rm peak}$ we equated the two expressions
for $L_{\rm iso}$.

\vskip 0.3 cm
\noindent
{\bf Radiative case --}
In this case an important fraction of the dissipated energy
is radiated away. 
This implies that the emitters, i.e. the electrons,
receive a large fraction of the available energy 
(directly or through the interactions with protons, and/or
through reconnection of the magnetic field) and
radiate it efficiently.
In this case the energy of the fireball decreases, changing
the $\Gamma(t)$ function.
This has been studied by Blandford \& McKee (1976); and the solution 
is (Katz \& Pian 1997; Vietri 1997; Sari, Piran \& Narayan 1998):
\begin{eqnarray}
\Gamma \, &=&\, { (\Gamma_0+1) (X+1)^2+(\Gamma_0-1) \over
                (\Gamma_0+1) (X+1)^2-(\Gamma_0-1)}
\nonumber  \\
X\, &=& \, {m \over M_{\rm f}}\, =\,  
{4\pi \Gamma_0 m_{\rm p} n c^2  R^3 \over  3E_{\rm k,f}  }
\label{grad}
\end{eqnarray}
where $M_{\rm f}$ is the mass of the fireball and $m$ 
is the swept interstellar mass.
When the fireball is decelerating, but still relativistic, 
$X\ll 1$ and
Eq. \ref{grad} simplifies to:
\begin{equation}
\Gamma  \sim  {1 \over X}  =  \left[ { 3E_{\rm k,f} \over
32\pi \Gamma_0 m_{\rm p} n c^5  a^3 t^3}\right]^{1/7} 
 =\,  \Gamma_0\, { t_{\rm dec}^{3/7} \over a^{3/7}}\, t^{-3/7}  
\label{gtrad}
\end{equation}
Inserting this into Eq. \ref{liso2} we obtain:
\begin{eqnarray}
L_{\rm iso,r} \, &=&\, 16 \pi  t^2 \Gamma_0^8  m_{\rm p} c^5 
\epsilon_{\rm e} n;
\quad \quad t\ll t_{\rm peak} 
\nonumber  \\
L_{\rm iso,r} \, &=&\,  
{3 \epsilon_{\rm e} E_{\rm k,f} \over 2 a^{10/7} }
t_{\rm dec}^{3/7}\, t^{-10/7}
\quad \quad\,\, t\gg t_{\rm peak} 
\nonumber  \\
t_{\rm peak,r} \, &=&\, {t_{\rm dec}\over a^{5/12}} 
\label{liso4}
\end{eqnarray}
The peak time of the bolometric afterglow emission 
(estimated equating the two limiting forms of $L_{\rm iso}$)
precedes the deceleration time by a small factor.
Integrating  $dR =2c \Gamma^2 dt$ assuming $\Gamma\propto t^{-3/8}$
(adiabatic) or $\Gamma\propto t^{-3/7}$ (radiative)
we have $a=4$ or $a=7$ for the adiabatic and radiative case,
respectively.
Therefore $t_{\rm peak} = 0.63 t_{\rm dec}$ (adiabatic)
and  $t_{\rm peak} = 0.44 t_{\rm dec}$ (radiative).

After the peak time, radiative afterglows decrease faster
than adiabatic ones, as the fireball energy is no longer 
constant, but decreases.
As noted by Sari, Piran \& Narayan (1998), partially
radiative fireballs would have scalings intermediate
between the pure adiabatic and pure radiative limits.
Even if, initially, a fireball is purely radiative,
after some time it must become adiabatic, as a consequence
of incomplete cooling of the accelerated electrons.
If the electrons are accelerated above some minimum
energy $\gamma_{\rm m} m_{\rm e}c^2$, this will occur when 
this electrons cannot cool in a dynamical time,
so when $\gamma_{\rm m} = \gamma_{\rm c}$, where
$\gamma_{\rm c}m_{\rm e} c^2$ is the energy
of those electrons cooling in $t^\prime\sim \Gamma t \sim R/(a c\Gamma) $.

When observing the flux in a particular frequency range $\Delta \nu$, 
we are never observing the bolometric flux, so in general the time decays
are different from $t^{-1}$ (adiabatic) or $t^{-10/7}$ (radiative).
If the emitted spectrum (in a $\nu F_\nu$ plot) has a peak at 
$\nu_{\rm peak}$, and $\nu_{\rm m}$ decreases in time,
then the time decay would be flatter for $\nu <\nu_{\rm peak}$,
and steeper for $\nu>\nu_{\rm peak}$.
However, if the observed flux has a spectral index
close to unity (i.e. $\nu\sim\nu_{\rm peak}$),
then the observed flux becomes a good proxy for the 
bolometric one, with the same time decay slope.

For a uniform circum--burst medium, the relation between 
the decay slope $\alpha$ and the spectral index $\beta$ for a flux 
density $F(\nu, t) \propto t^{-\alpha} \nu^{-\beta}$ is
(Sari, Piran \& Narayan 1998):
\begin{equation}
\alpha \, =\, {2\over 7}(6\beta-1)
\label{closure1}
\end{equation}
returning $\alpha=10/7=1.43$ when $\beta=1$ and
$\alpha=1.77$ for $\beta=1.2$.
This derivation assumes that the number of accelerated electrons
is always a fixed fraction of the protons present in the
circum--burst medium.

\section{Pair--enriched interstellar medium}

When the prompt phase emission spectrum extends above
$E_{\rm peak}(1+z)\sim m_{\rm e}c^2$ we can convert
a fraction of the high energy photons into electrons--positron pairs.
This case has been studied in detail by Thompson \& Madau (2000),
Meszaros, Ramirez--Ruiz \& Rees (2001) and especially by 
Beloborodov (2002).

The basic idea is that although the scattering depth of the
circumburst medium can be much smaller than unity,
it can nevertheless scatter a fraction of the prompt phase
photons along non radial directions. 
These scattered photons can then interact with the arriving 
high energy prompt phase photons producing pairs.
The process {\it is not} controlled by the probability of the
interaction between the scattered and the primary prompt phase photons:
this is almost unity (up to very large distances), due to the 
huge amount of the prompt phase photons.
The process is controlled by how many photons are scattered.
The full description of this scenario is rather complex, and
we refer to Beloborodov (2002) for the complete treatment.
We focus here on a few estimates, to give the idea of the importance
of the process.
The basic quantity of interest is the number of scatterings
done by a single electron located at a distance $R$ from the
emission site of the prompt phase emission.
Using the Thomson cross section for simplicity, and
setting $h\nu \equiv x m_{\rm e} c^2$,  this number is
\begin{equation}
N_{\rm sc} \, =\, \sigma_{\rm T} \, 
{E_{\rm \gamma,iso} \over \langle x\rangle m_{\rm e} c^2 4\pi R^2 
ct_{\rm burst}} \, ct_{\rm burst} \, \sim \, 640 \, 
{E_{\rm \gamma,iso,54} \over \langle x\rangle R_{16}^2 }
\label{pairs}
\end{equation}
Almost all these photons will be converted into pairs immediately
after they have been scattered.
This implies that the  circumburst medium will be greatly enriched
by pairs before the arrival of the forward shock.
This can occur even if the total number of the intercepted photons
is a tiny fraction of the total. 
For instance, if the interstellar medium is homogeneous 
with density $n$, the total number of scattered photons
within $10^{17}$ cm is only a fraction $\tau_{\rm T} = 6.65\times 10^{-8}n$
of the total number of photons of the prompt phase.
But this is enough to greatly pair--enrich the circumburst medium.  
Furthermore, the scattering and the pair production processes 
pre--accelerate the interstellar medium. 
If there is one proton per primary electron, and if the 
energy deposited by the single scattering with subsequent pair
production is roughly equal to $m_{\rm e} c^2$, this process
will be important below a certain distance, below which 
there occur more than 1000 scatterings for primary electron
(i.e. in this case the proton associated with the primary electron
will start to move with $\Gamma\sim 2$ in the radial direction).
As a feedback, if the medium starts to move then the
typical energy of the scattered photons will start to decrease,
quenching off the pair production process (i.e. the scattered
photons have too small energies to interact with photons around a few MeV).
On the other hand, the produced pairs, if are re--isotropized in a short
time, can also scatter the incoming prompt phase radiation, enhancing the
process and making it exponential.

Therefore Eq. \ref{pairs} is only a simple but rough estimate of 
a much more complex scenario.
We can nevertheless draw some important conclusions:
\begin{itemize}
\item 
Pairs are important if the prompt phase emission extends
above threshold.
\item 
At a negligible expense (i.e. the fraction of absorbed prompt phase
emission is negligible) the environment is largely enriched by pairs.
\item 
The distance for which the number of produced pairs equals
the number of primary electrons is sufficiently large and affects
the properties of the forward shock up to some relevant
observed time.
For instance, the ``closure" relation given by Eq. \ref{closure1} 
is modified as long as the number of pairs per proton
is larger than unity, because in this case
the energy $\gamma_{\rm m}\propto \Gamma n/n_+ \propto \Gamma R^2$.
Here $n_+$ is the pair density.
Introducing this extra $R^2$ dependence we find
\begin{equation}
\alpha \, =\, {2\over 7}(4\beta+1) 
\label{closure2}
\end{equation}
returning $\alpha=10/7=1.43$ when $\beta=1$ and
$\alpha=1.66$ for $\beta=1.2$.

\item Although the details of the shock acceleration process are
controversial, it is reasonable to assume that the ratio of the 
energy given to leptons and protons will increase, if we have
many leptons per proton.
This is then one way to have a radiative fireball.
\end{itemize}
We therefore propose that bursts whose prompt phase
emission extends above $m_{\rm e} c^2$ should be characterised 
by an early radiative (then powerful) afterglow.

\subsection{Additional processes}

\begin{figure}
\vskip -0.3 cm
\hskip -1.6 cm
\psfig{figure=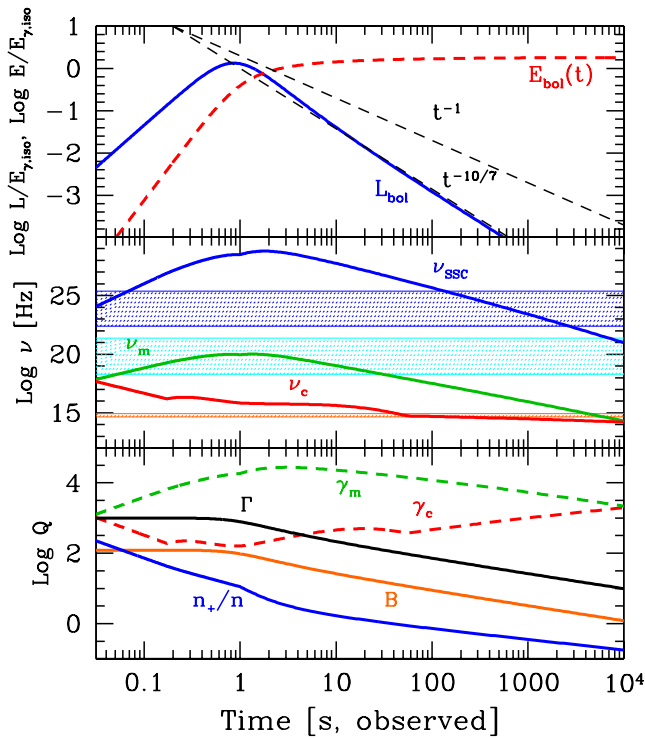,width=11.5cm}
\vskip -0.8 cm
\caption{
Top panel: time profiles of the bolometric
luminosities and the corresponding cumulative energetics,
in units of the initial kinetic energy of the fireball.
For this particular example, we have assumed a radiative
fireball with $z=1$, $E_{\rm \gamma,iso}=10^{53}$ erg, $\eta=0.2$,
$T_{90}=1$s, $\Gamma_0=10^3$, $\epsilon_{\rm e}=0.9$ and $p=2$.
The circumburst medium is homogeneous with density $n=1$ cm$^{-3}$.
The dashed lines corresponds to $t^{-1}$ and $t^{-10/7}$,
i.e. the adiabatic and radiative cases.
Pair production is accounted for in a approximated way, assuming that
all scattered photons are transformed into pairs, but assuming
that there are at most $m_{\rm p}/m_{\rm e}$ pairs per primary electron.
Mid panel: the time profiles of the frequencies
$\nu_{\rm m}$, $\nu_{\rm c}$ and 
$\nu_{\rm SSC}\equiv \gamma_{\rm m}^2\nu_{\rm m}$.
The hatched areas mark the energy ranges of the LAT instrument
[0.1--100 GeV], the GBM instrument [8--1000 keV] and the optical range 
(corresponding to the $U$ and $R$ filters).
Bottom panel: the time profiles of the injected energy
$\gamma_{\rm m}$ and  the cooling energy $\gamma_{\rm c}$.
We also show the profile of $\Gamma$, of the magnetic
field $B$ (assuming $\epsilon_{\rm B}=0.1$), and
the number of pairs per primary electron $n_+/n$.
Since $n=1$, this also corresponds to the density of pairs.
}
\label{example}
\end{figure}

We consider here other processes that can be relevant for
the formation of the high energy afterglow:

\begin{itemize}

\item When $t_{\rm bursts}>t_{\rm dec}$
the region of the forward shock where leptons are accelerated
is illuminated by the flux of the prompt phase emission
(of luminosity $L_{\rm \gamma,iso}$).
This component lasts as long as the forward shock is 
illuminated by the prompt phase 
(see Beloborodov 2005a).
The corresponding energy density, as measured in the 
comoving frame of the forward shock is
\begin{equation}
U^\prime_{\rm ext} \, =\, { L_{\rm \gamma,iso} \over
4\pi R^2 c \Gamma^2} 
\end{equation}
where the subscript ``ext" stands for ``external" to the 
afterglow emitting region.
This has to be compared with the local magnetic energy density
\begin{equation}
U^\prime_{\rm B} \, =\, \epsilon_{\rm B} n m_{\rm p}c^2 \Gamma^2
\end{equation}
Therefore the ratio between the synchrotron and the ``external Compton"
(i.e. the luminosity produced by scattering $U^\prime_{\rm ext}$)
luminosities is 
(see also Beloborodov 2005a):
\begin{equation}
{L_{\rm EC} \over L_{\rm syn}}  = 
f\, {U^\prime_{\rm ext} \over U^\prime_{\rm B} } = 
{f\,  L_{\rm \gamma,iso} \over 
4\pi R^2\Gamma^4   \epsilon_{\rm B} n m_{\rm p}c^3 }
 = 0.18 \, 
 { f\, L_{\rm \gamma,iso,53} \over 
 R^2_{17}\Gamma_3^4  \epsilon_{\rm B,-1} n  }
\end{equation}
The factor $f<1$ accounts for the suppression  
of the power emitted in the direction of the observer
due to the anisotropic pattern of the incoming photons in
the frame of the fireball.
An order of magnitude estimate of its value can be gained
through a simple example.
In the frame of the fireball, assume that all the seed photons
for the scattering are coming radially.
Electrons travelling at $\theta^\prime=180^\circ$ from the photons 
lose energy at a rate $\propto \gamma^2(1-\beta\cos\theta^\prime) \sim 4\gamma^2$.
Electrons moving at $90^\circ$ lose energy at a rate $\propto \gamma^2$.
This is the emission that the observer (on the Earth) will 
preferentially see. Therefore the factor $f$ is less than, but of order of,
unity.
This external Compton component would start to be important
at frequencies above $\nu_{EC}\sim \gamma_{\rm m}^2 \nu_{\rm peak}\sim 
\gamma_{\rm m,3}^2 \nu_{\rm peak,MeV}$ TeV.
Below $\nu_{\rm EC}$ we should have $F(\nu)\propto \nu^{-1/2}$.

\item
The high energy emission can also be produced by the synchrotron
self--Compton (SSC) process
(See e.g. Corsi et al. 2009; Fan et al. 2008), particularly important when
i) $\epsilon_{\rm e}>\epsilon_{\rm B}$; 
ii) we are in the fast cooling regime and 
iii) we are in the Thomson limit (i.e. the scattering can be described 
by the Thomson cross section).

Condition i) and ii) are always fulfilled in radiative fireballs,
while condition iii) may be violated. 
The limit for the Thomson regime can be derived considering
the dimensionless frequency $x^\prime_{\rm m} = 
h \nu_{\rm m}/(\Gamma m_{\rm e}c^2)$ (as measured in the comoving frame) 
and the electron energy $\gamma_{\rm m}$.
If $x^\prime_{\rm m} \gamma_{\rm m}>1$ 
The entire process occurs in the Klein Nishina regime
if $x^\prime_{\rm m} \gamma_{\rm m}>1$, i.e. when: 
\begin{equation}
\Gamma^3 \epsilon_{\rm e}^3 \epsilon_{\rm B}^{1/2} n^{1/2}
\left[ { m_{\rm p}\over m_{\rm e} } \, {n\over n_+} \right]^3 \, >
1.77 \times 10^{14}
\end{equation}
For moderate pair production (i.e. $n_+/n \lsim 100$) and
for still large $\Gamma$ the early SSC process is then 
in the Klein Nishina regime, and is therefore inefficient.
Furthermore, the SSC spectrum starts
to be important at $\nu_{\rm SSC}$ given by
\begin{equation}
\nu_{\rm SSC}\, \sim \, \gamma_{\rm m}^2\nu_{\rm m} \, \sim\,
7\times 10^{22} \, \epsilon_{\rm e}^4 \epsilon_{\rm B}^{1/2} n^{1/2}
\Gamma_3^6 \left[{ m_{\rm p}\over m_{\rm e} } \, {n\over n_+}  \right]^4
\,\,\, {\rm Hz}
\end{equation}
It is a strong function of $n/n_+$: for less than 100 pairs per proton
(and still a large $\Gamma$) the SSC spectrum starts at frequencies
above the LAT range (with a flux reduced by Klein---Nishina effects).
The mid panel of Fig. \ref{example} shows $\nu_{\rm SSC}$ as a function of time
for one particular case.
\end{itemize}

We conclude that the most likely radiation process originating
the LAT emission is synchrotron. 

To illustrate the above considerations and to give an example of the 
predicted high energy flux in radiative fireballs,
we have calculated the bolometric flux emitted in one specific case,
assuming that the prompt phase energetics 
$E_{\rm \gamma, iso}=10^{53}$ ergs, $\eta=0.2$, $z=1$, 
$\Gamma_0=1000$, $n=1$ cm$^{-3}$, $p=2$.
Furthermore, we assumed a duration of 1 s
and $\epsilon_{\rm e}=0.9$, $\epsilon_{\rm B}=0.1$.
The resulting bolometric luminosity (normalised to $E_{\rm \gamma,iso}$) 
is shown in the top panel of Fig. \ref{example}, together with its
corresponding energetics [$E_{\rm bol}(t) =\int_0^t L_{\rm bol}(t')dt'$].
We also indicate the $t^{-1}$ and the $t^{-10/7}$ time behaviour
(dashed black lines).
The mid panel shows the time profile of 3 characteristic frequencies:
the injected frequency $\nu_{\rm m}$, the cooling frequency
$\nu_{\rm c}$ and the SSC frequency 
$\nu_{\rm SSC}\equiv \gamma_{\rm m}^2\nu_{\rm m}$
(see also Beloborodov 2005b for the case of pair enriched circum--burst
material, but with an adiabatic fireball).

The 2 upper shaded areas correspond to the frequency ranges covered by the LAT
and GBM, while the lower one indicates the optical frequency range.
The bottom panel shows the time profile of the minimum Lorentz factor 
of the injected electron $\gamma_{\rm m}$, the cooling Lorentz factor
$\gamma_{\rm c}$, the bulk Lorentz $\Gamma$, together with the
time profile of the magnetic field $B$ and the number of pairs
per proton $n_+/n$, calculated according to Eq. \ref{pairs}.
This quantity is crucial to calculate $\gamma_{\rm m}$,
since the same available energy must be divided by the
total number of leptons, including the pairs.
Since their amount changes with $R$ (and correspondingly
with the observed time), the time profile of $\gamma_{\rm m}$
is greatly modified by the presence of pairs.
As a consequence, both $\nu_{\rm m}$ and $\nu_{\rm SSC}$ are
largely affected, their values being much lower than 
in the absence of pairs.
A note a caution: although the presence of pairs
may be crucial to bring the process to the radiative
regime, the exact amount of pairs is difficult
to calculate, being partially dependent on the 
exact shape and time evolution of the
spectrum of the prompt phase emission above threshold, 
the presence or not of a magnetic field embedded in the 
circum--burst medium, a possible clumping of this medium, and so on.
Ours are bound to be only rough estimates.
Bearing the above caveat in mind, we find that the
synchrotron emission, at the peak time, should have a flux
$F(\nu)\propto \nu^{-0.5}$ between $\nu_{\rm c}$ and $\nu_{\rm m}$
and $F(\nu)\propto \nu^{-p/2}$ (equal to $\nu^{-1}$ in this example)
up to $\nu_{\rm max}=\nu_{\rm m} (\gamma_{\rm max}/\gamma_{\rm m})^2$.
Therefore $\gamma_{\rm max}/\gamma_{\rm m}\gsim 10^3$ ensures that the
synchrotron emission extend up to the GeV range.

Note that the $\nu^{-p/2}$ part of the spectrum may start
in the GBM energy range, depending on the exact amount of pairs.
There is then the possibility that the afterglow emission
``contaminates" the prompt phase emission seen by the GBM.
In some cases, this ``contamination" can appear as an excess 
at both extremes of the GRB energy range, especially if pairs
are very important, decreasing $\gamma_{\rm m}$ (as in the case of GRB 090902B,
Abdo et al. 2009c).
Also the opposite (i.e. the prompt phase ``contaminates" the afterglow
seen in the LAT) can occur, especially
when the high energy Band index $\beta$ is not too soft.
In this latter case most of the prompt phase photons contributing
to the LAT flux should be at low energies.

For simplicity, we have assumed that $\epsilon_{\rm e}$ is
constant, and not proportional to the amount of pairs per proton
(since this number is uncertain).
However, the radiative phase should end in any case when 
$\gamma_{\rm c}$ becomes greater than $\gamma_{\rm m}$ since in this
case most of the energy given to electrons cannot be radiated away in a
dynamical time.

\begin{figure}
\vskip -0.7 cm
\hskip -0.6 cm
\psfig{figure=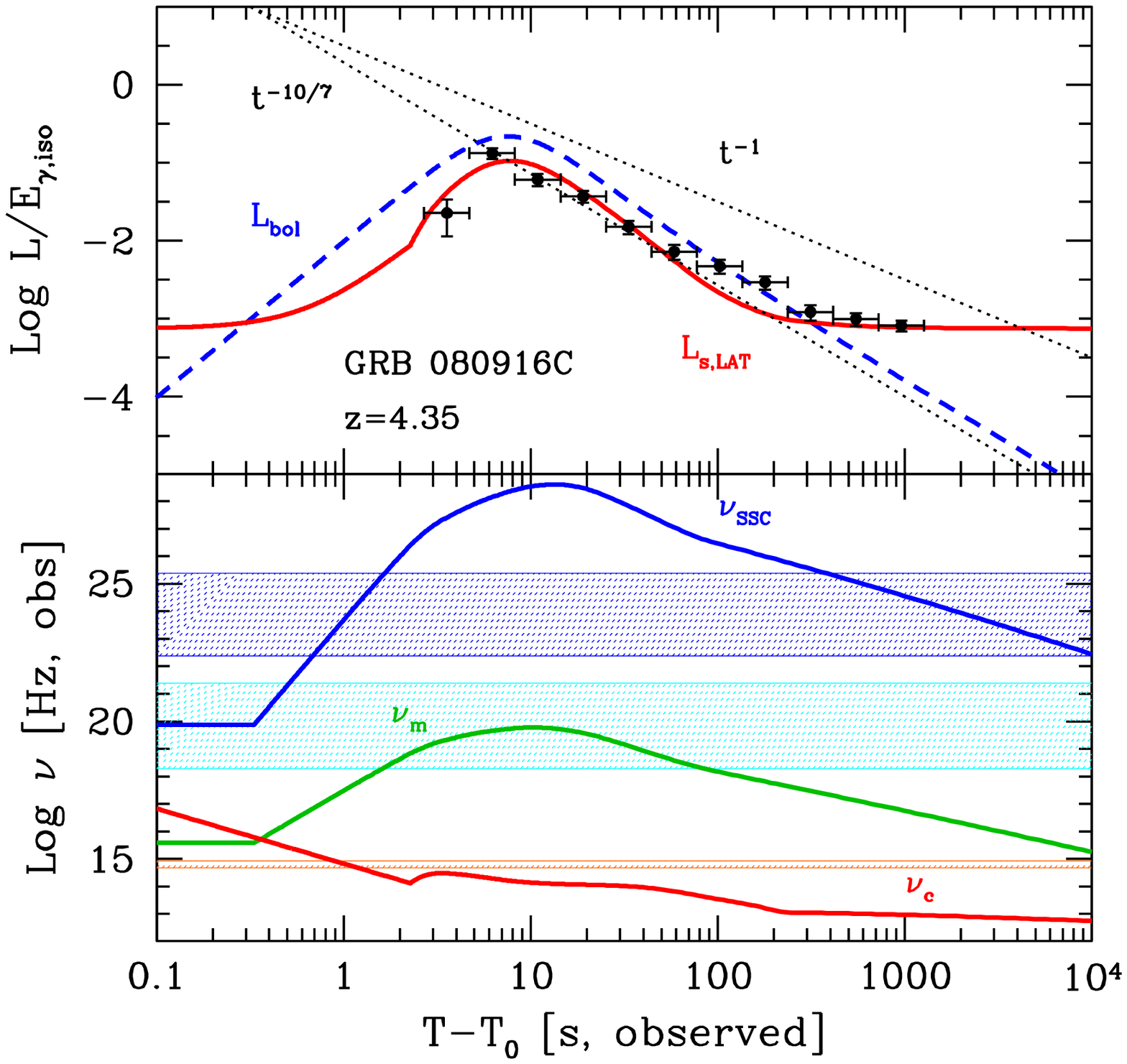,width=9.5cm,height=11cm}
\vskip -0.8 cm
\caption{The long burst GRB 080916C.
Parameters are listed in Tab. \ref{para}.
}
\label{f080916c}
\end{figure}

\begin{figure}
\vskip -0.7 cm
\hskip -0.6 cm
\psfig{figure=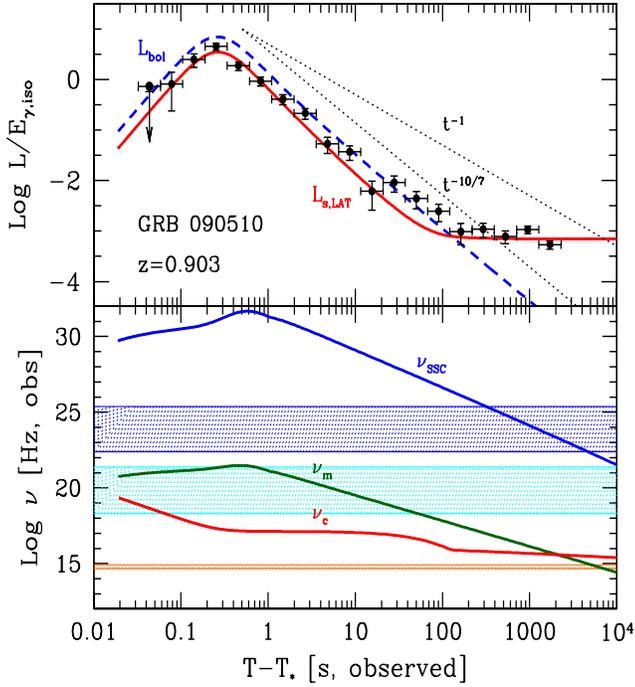,width=9.5cm,height=11cm}
\vskip -0.8 cm
\caption{The short burst GRB 090510 assuming 
$T_*=0.6$ s.
Parameters are listed in Tab. \ref{para}.
}
\label{f090510}
\end{figure}

\begin{figure}
\vskip -0.7 cm
\hskip -0.6 cm
\psfig{figure=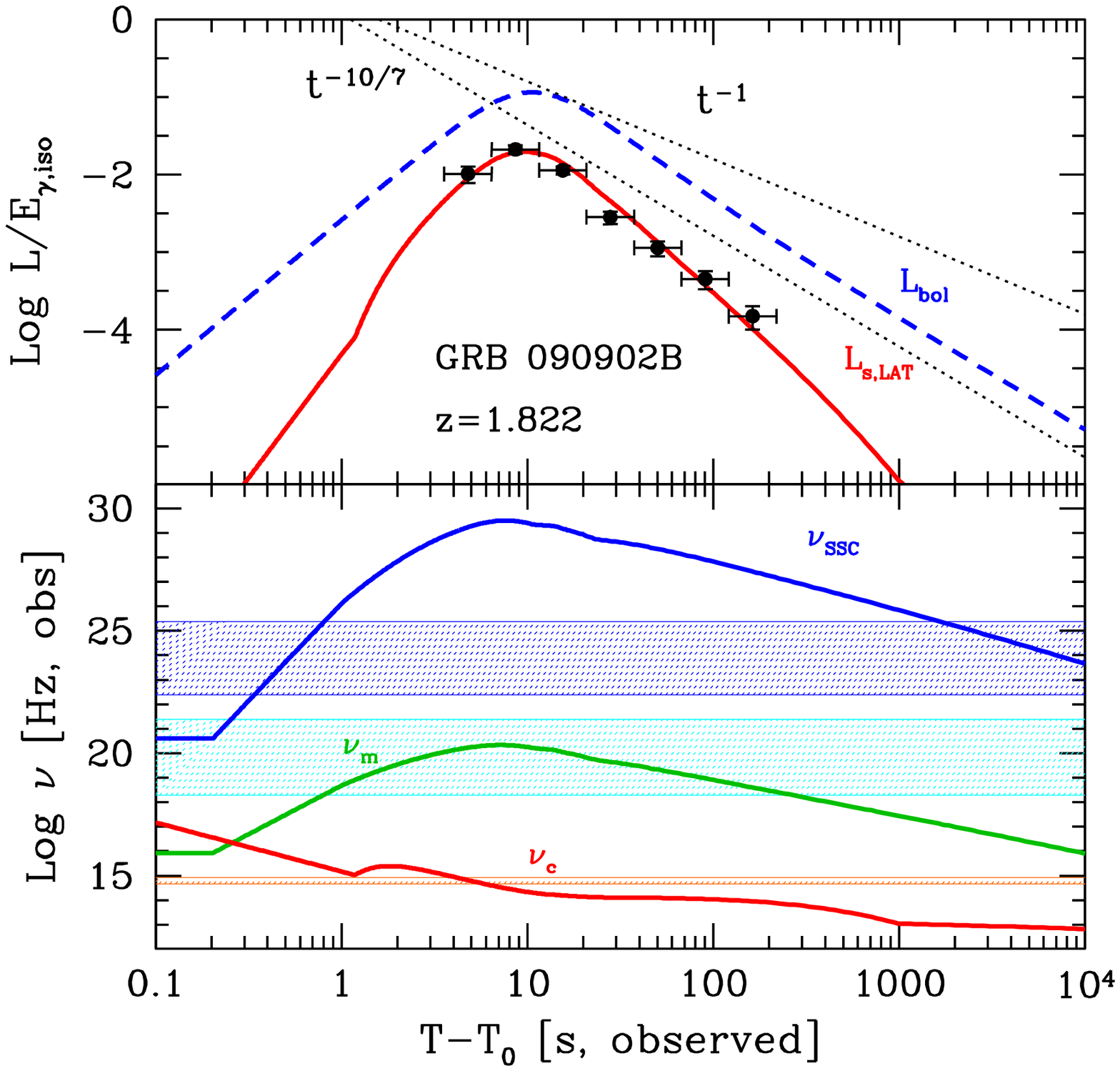,width=9.5cm,height=11cm}
\vskip -0.8 cm
\caption{The long burst GRB 090902.
Parameters are listed in Tab. \ref{para}.
}
\label{f090902}
\end{figure}

\begin{figure}
\vskip -0.7 cm
\hskip -0.6 cm
\psfig{figure=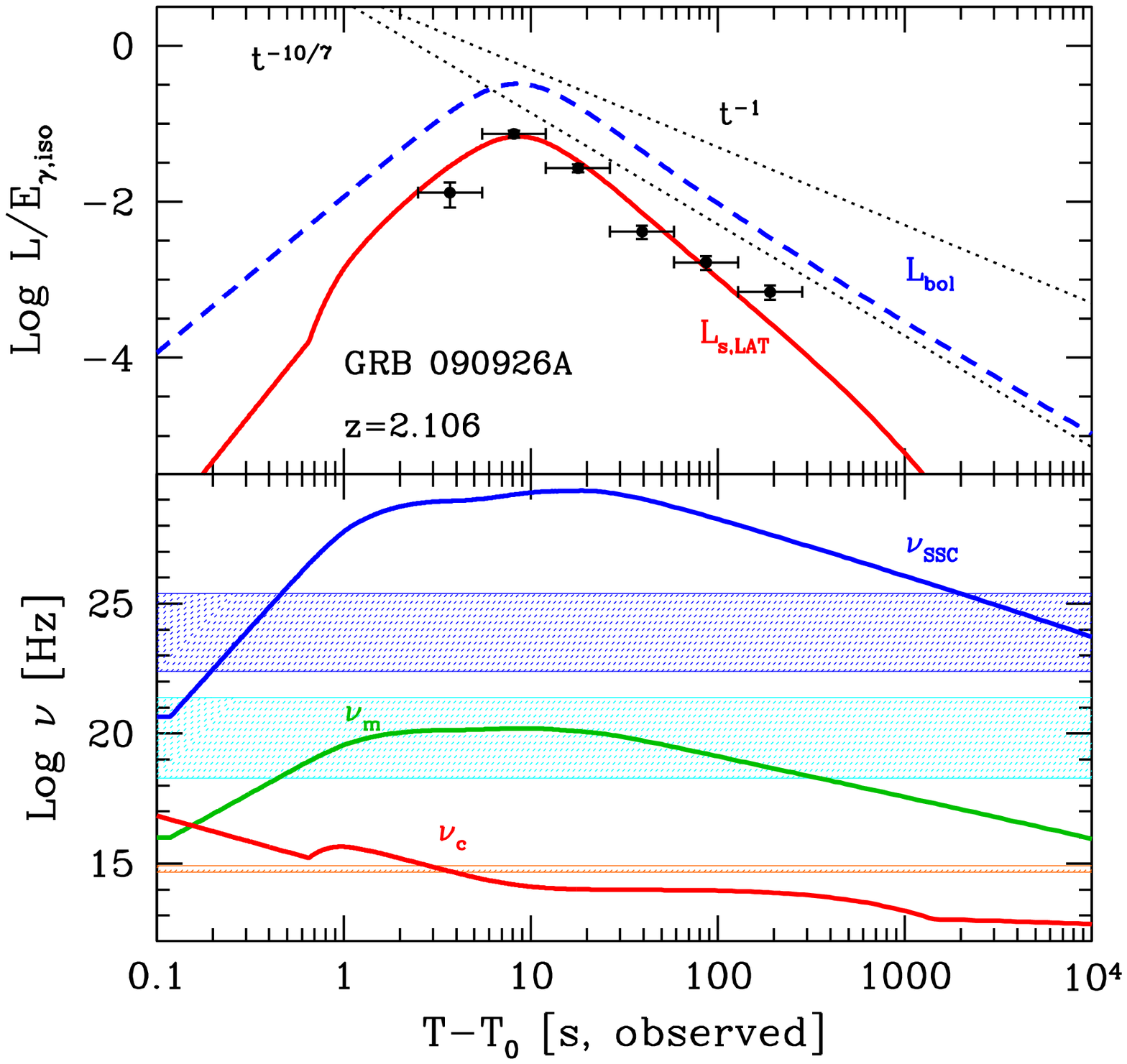,width=9.5cm,height=11cm}
\vskip -0.8 cm
\caption{The long burst GRB 090926.).
Parameters are listed in Tab. \ref{para}.
}
\label{f090926}
\end{figure}

\begin{table}
\caption{Parameters for the radiative afterglow models}
\label{tab2}      
\centering  
\begin{tabular}{l l l l l l}       
\hline
\hline       
GRB         
              &$\Gamma_0$ &$E_{\rm \gamma,iso}$ &$\eta$ &$n$        &$p$  \\
              &           & &       &cm$^{-3}$  &     \\ 
\hline
Fig. \ref{example}
              &1000 &1.0e53 &0.2  &1   &2    \\
080916C       &900  &5.6e54 &0.32 &2   &2   \\
090510        &2000 &5.0e52 &0.13 &0.1 &2.1  \\
090902B       &630  &4.4e54 &0.25 &2   &2.6   \\	
090926        &670  &2.0e54 &0.14 &3   &2.5   \\	
\hline
\hline                  
\end{tabular}
\label{para}
\end{table}

\section{Application to specific bursts}

We applied the radiative scenario to the 4 brightest (in the LAT) 
GRBs with redshift.
They are the same illustrated in Fig. \ref{f2}, namely
GRB 080916C, GRB 090510, GRB 090902B and GRB 090926.
In principle, the number of parameters used for the
adopted model is limited (they are listed in Tab. \ref{para}), 
but we adopted a few rather drastic simplifications:
\begin{itemize}
\item
We consider the fireball, when colliding with the
interstellar medium, as ``thin".
In other words, we assume that it can act as a piston
having a total energy $E_{\rm k,f}$.
This is completely right for short GRBs, but not for long ones.
According to Fig. \ref{lc}, the {\it Fermi}/LAT emission
of several GRBs starts while the emission seen by the GBM has not ended.
In this case the $t^2$ rising behaviour of the LAT light curve
can be different (see Sari 1997).

\item 
When calculating the number of pairs produced
by the circum--bursts medium, we neglect the 
amplification (exponential) effect of the produced
pairs that can themselves scatter the incoming radiation.
The momentum deposited in the circum--bursts medium
is also taken into account only by imposing that the maximum
number of pairs per proton is $m_{\rm p}/m_{\rm e}$,
since a larger number corresponds to a mildly relativistic motion
of the medium, and the quenching off of the pair--producing mechanism
For simplicity, we use the Thomson cross section for scattering,
and assume that most of the prompt phase photons are close to the
threshold for pair production.

\item
We assume that {\it all} electrons and positrons 
are accelerated.
If, instead, only a fraction of them receive the entire
available energy, then the typical Lorentz factors of the
accelerated leptons is larger.

\item
We use a fixed value of $\epsilon_{\rm e}$, even if the number
of pairs populating the circum--bursts medium decreases with $R$.
Consequently, we use the radiative solution all throughout the
shown evolution, with no transition to the adiabatic case.
\end{itemize}

Bearing in mind these caveats, Figs. \ref{f080916c}--\ref{f090926}
show the light--curves of the 4 GRBs interpreted on the basis
of our radiative model, with the main parameters listed in 
Tab. \ref{para}.
In the cases of GRB 080916C and GRB 090510
we have also added a constant flux to the light curve,
to account for the presence of the background,
flattening off the observed light--curves.
In the case of GRB 090510, the fact that the flux above 200 s is 
due to the background has been confirmed by De Pasquale et al. (2009,
see their Fig. 1).
Also for 080916c the points above 1000 s are affected by
background (see Abdo et al. 2009a, and their Fig. 4).
So, for these two bursts, the flattening of the light--curve
at late times should not be due to the contribution of the SSC component 
entering in the LAT energy range (as predicted by Dermer,
Chiang \& Mitman 2000, and tantalisingly suggested
by Fig. \ref{f080916c} and Fig. \ref{f090510}), 
but only because we did not subtract the background.

The solid lines shown in all top panels refer to the 
luminosity integrated in the 0.1--100 GeV energy range,
while the dashed thick lines are  the bolometric fluxes
(both normalised to the prompt phase energetics of each burst).
For comparison we show also the lines corresponding to
$t^{-1}$ and to $t^{-10/7}$.
We can see that in all cases the radiative interpretation 
is in good agreement with what observed, and that
in all cases the predicted $\nu_{\rm m}$ is well below 
the 0.1 GeV value.
This ensures that in the LAT we should see a spectral shape
$F(\nu) \propto \nu^{-p/2}$.
The observed decay slope
and the spectral index in the LAT energy range (see Tab. 1)
are consistent with Eq. \ref{closure2},
but the errors on $\beta=\Gamma_{\rm LAT}-1$
are too large to use this as a reliable test.

\section{Discussion}

The found bulk Lorentz factors are in the range 630--900
for the long bursts, and 2000 for the short GRB 090510.
We believe that these relatively large values are the key
to understand why only a minority of bursts
are detectable by the LAT.
A large bulk Lorentz factor, in fact, means an early
peak time of the afterglow (see Eq. \ref{liso3} and Eq. \ref{liso4}),
and this in turn means a large flux.
{\it Faster fireballs have brighter afterglows}.
This is true for adiabatic as well as radiative fireballs.
If the emission occurs in the radiative regime then the afterglow 
will be brighter still, since all the energy dissipated in the external 
shock is radiated away.

If the circum--bursts medium is enriched by electron--positron pairs,
we have a more favourable set up for a radiative process.
If the acceleration mechanism divides its energy to all particles,
then leptons should receive a total energy
exceeding the one given to protons.
But this may be only one of the means to have a radiative fireball.
An alternative is to have a strong coupling between electrons and
protons, with an efficient energy flow from protons to electrons.
In any case, we can easily test if pairs are indeed important
by simply comparing the general properties of the early afterglow
for bursts of different $E_{\rm peak}$ and high energy index $\beta$,
since only those bursts whose prompt phase photon energies exceed $m_{\rm e}c^2$
should efficiently populate the circumburst medium by pairs.
As an example, we may test if the high energy emission is present only
in GRBs of high $E_{\rm peak}$ (in the rest frame) as it appears to be
the case until now, or if it occurs also for bursts with a small $E_{\rm peak}$.
If this will occur, and if the flux will decay with a slower rate than $t^{-10/7}$,
then we will have an indication of a fast fireball that emits adiabatically because of
no pairs--enrichment of the circum--bursts medium.
In other words, a possible test of the idea of having radiative afterglows
because of pair enrichment is to find a different time decay for
the high energy emission in classical GRBs whose prompt phase emission extends 
to high energies and
X--ray flashes, characterised by relatively small values of $E_{\rm peak}$.

The radiative interpretation could ease the efficiency problem of the afterglow phase.
This problem concerns the ratio of the energetics emitted during the 
prompt and afterglow phases, that is much larger than unity
(e.g. Zhang et al. 2007).
According to the standard internal/external shock scenario one expects the opposite,
since external shocks should be much more efficient than internal ones
to dissipate the kinetic energy of the fireball.
These estimates were based on the observed X--ray afterglow energetics
(see e.g. Willingale et al. 2007; Ghisellini et al. 2009), 
and we can now revise them including the much more powerful
high energy $\gamma$--ray emission, bringing the total afterglow energetics
to be roughly equal to the prompt phase one.
Furthermore, if the fireball is indeed radiative in the first phases,
with a consequent fast decay, we can understand why the afterglow emission
at later times and at other frequencies is so faint.

According to our findings, bursts detected by the LAT may be the ones
with the largest $\Gamma$, and can be used to explore the high--end
$\Gamma$--distribution.
On the other hand, one can wonder about the possibility to detect
with the LAT bursts with relatively smaller $\Gamma$, smaller
high energy luminosities and with light curves peaking at larger
peak times. 
Even if rare, nearby objects with these properties might be still detectable, 
offering a direct way to test our ideas: even if they should be characterised 
by much lower peak luminosities in the LAT, they should have LAT/GBM fluence 
ratios similar to those presented in this paper, and lower values of $\Gamma$.

One of the argument put forward against the afterglow interpretation
of the high energy flux is its variability, that according to
Abdo et al. (2009c) can have a timescales $\Delta t_{\rm var}$ as short as 90 ms.
If true, this is certainly a severe problem for the afterglow interpretation.
On the other hand the knowledge of $\Delta t_{\rm var}$ 
is limited by the few number of received photons.
When the entire light--curve, lasting for a few hundreds seconds,
is composed by a few hundreds events, one can define a very short $\Delta t_{\rm var}$ 
only if there is an exceptional ``bunching" of photons
in contiguous time--bins, and we do not see it in the bursts we analysed.

Finally, we would like to emphasise the importance of establishing, in general,
if the high and low energy emission are produced  by the same electrons
at the same time or instead if they are produced by different electrons
at different times.
As the study of GRB 090510 (Abdo et al. 2009b; Ghirlanda et al. 2009) 
has demonstrated, we are reaching the required data quality
to put strong constraints on the theories predicting the violation of the Lorentz
invariance at small scales, that can be tested by comparing the possible delay 
of the {\it arrival} times of high energy photons.
The critical issue about these studies is to know exactly the {\it generation} time
of the high with respect to low energy emission.
Therefore it becomes crucial to establish if the flux received by the LAT is the extension
in energy of the prompt phase emission or if it is afterglow radiation.

\section*{Acknowledgments}
We thank the anonymous referee for useful suggestions.
This work was partly supported by a 2007 COFIN--MIUR grant and by ASI grant 
I/088/06/0. This work is based on the publicly available \fe\ data obtained through the 
Science Support Center (SSC).  
ASI--ASDC is acknowledged for useful tutorials on the 
\fe\ data analysis. We thank F. Tavecchio for useful discussions and suggestions.

\end{document}